\documentclass[conference,compsoc]{IEEEtran}
\usepackage[utf8]{inputenc}
\usepackage[T1]{fontenc}

\ifCLASSOPTIONcompsoc
\usepackage[nocompress]{cite}
\else
\usepackage{cite}
\fi
\ifCLASSINFOpdf
\usepackage[pdftex]{graphicx}
\DeclareGraphicsExtensions{.pdf,.jpeg,.png}
\else
\fi
\usepackage{amsmath}
\usepackage{amsthm}
\usepackage{amssymb}
\usepackage{mathtools}

\usepackage{algorithm}
\usepackage[noend]{algpseudocode}

\algnewcommand\algorithmicswitch{\textbf{switch}}
\algnewcommand\algorithmiccase{\textbf{case}}
\algnewcommand\algorithmicassert{\texttt{assert}}
\algnewcommand\Assert[1]{\State \algorithmicassert(#1)}%
\algdef{SE}[SWITCH]{Switch}{EndSwitch}[1]{\algorithmicswitch\ #1\
  \algorithmicdo}{\algorithmicend\ \algorithmicswitch}%

\algdef{SE}[CASE]{Case}{EndCase}[1]{\algorithmiccase\
  #1}{\algorithmicend\ \algorithmiccase}%

\algtext*{EndSwitch}%
\algtext*{EndCase}%

\usepackage{array}

\usepackage{multirow}

\ifCLASSOPTIONcompsoc
\usepackage[caption=false,font=footnotesize,labelfont=sf,textfont=sf]{subfig}
\else
\usepackage[caption=false,font=footnotesize]{subfig}
\fi

\usepackage{stfloats}
\usepackage{url}

\usepackage{tikz}
\usetikzlibrary{shapes,arrows,automata,backgrounds,positioning,patterns,decorations.markings}

\usepackage{color}

\definecolor{myred}{RGB}{215,0,0}
\definecolor{mygreen}{RGB}{0,215,0}

\tikzstyle{noeud}=[rectangle, draw=black, rounded corners,
text centered,text width=1.35cm, rectangle split,
rectangle split horizontal=true, rectangle split parts=2,
rectangle split part fill={mygreen, myred}, font=\footnotesize]
\tikzstyle{myarrow}=[->,>=stealth',shorten >=1pt]

\tikzstyle{noeudex}=[rectangle, draw=black, rounded corners,
text centered,rectangle split,
rectangle split horizontal=true, rectangle split parts=2,
rectangle split part fill={mygreen, myred}, font=\footnotesize]

\tikzstyle{myrect}=[rectangle, draw,minimum height=0.5cm, text width=0.26cm]

\newcommand{\world}{\mathcal{W}}

\newcommand{\bad}{\mathcal{B}}

\newcommand{\node}[1][v]{\mathnormal{#1}}

\newcommand{\unsafe}{\mathcal{U}}

\newcommand{\init}{\mathcal{I}}

\newcommand{\badcube}{\gamma}

\newcommand{\te}[1]{\texttt{#1}}

\newtheorem{myrule}{Rule}

\begin{document}
%
\title{FAR-Cubicle - A new reachability algorithm for Cubicle}


\newcommand{\lri}{LRI, Universit\'e Paris Sud CNRS, Orsay F-91405}
\newcommand{\scl}{Intel Corporation}
\newcommand{\inria}{INRIA Saclay -- Ile-de-France, Orsay cedex, F-91893}
\newcommand{\mpi}{Max Planck Institute for Software Systems}
\newcommand{\apple}{Apple}

\author{
  \IEEEauthorblockN{
    Sylvain Conchon\IEEEauthorrefmark{1}\IEEEauthorrefmark{2} \qquad
    Amit Goel\IEEEauthorrefmark{3} \qquad
    Sava Krsti\'c\IEEEauthorrefmark{4} \qquad
    Rupak Majumdar\IEEEauthorrefmark{5} \qquad
    Mattias Roux\IEEEauthorrefmark{1} \qquad
  }
  \\
  \IEEEauthorblockA{\IEEEauthorrefmark{1}\lri}
  \IEEEauthorblockA{\IEEEauthorrefmark{2}\inria}
  \IEEEauthorblockA{\IEEEauthorrefmark{4}\scl}
  \IEEEauthorblockA{\IEEEauthorrefmark{4}\mpi}
  \IEEEauthorblockA{\IEEEauthorrefmark{3}\apple}
  
}


%


\maketitle

\begin{abstract}
  We present a fully automatic algorithm for verifying safety
  properties of parameterized software systems. This algorithm is based
  on both IC3 and Lazy Annotation. We implemented it in Cubicle, a model
  checker for verifying safety properties
  of array-based systems. Cache-coherence protocols and mutual exclusion
  algorithms are known examples of such systems. Our algorithm
  iteratively builds an abstract reachability graph refining the set of
  reachable states from counter-examples. Refining is made through
  counter-example approximation. We show the effectiveness and
  limitations of this algorithm and tradeoffs that results from
  it.
\end{abstract}


%
\IEEEpeerreviewmaketitle

\section{Introduction}
We describe FAR (Forward Abstracted Reachability), an algorithm for
fully automatic verification of parameterized software systems. A
parameterized system describes a family of programs such as cache
coherence protocols where the number of processes involved can change
but the algorithm handling their behaviour is the same for all of
them. Thus, the parameter allows to talk about these algorithms
without knowing the actual number of processes that will be involved
and then to prove its safety regardless of this number.

Safety properties state that ``nothing bad happens'' in our
parameterized system. Verifying them
can be reduced to finding an invariant of it. Finding an invariant can
be hard (and even undecidable \cite{Abdulla96}). The standard approach
to find one is to find a formula $\Phi$ such that $\Phi$ is an
inductive invariant of the system (\textit{i.e.} the initial state of
the system satisfies $\Phi$ and taking a transition from a state
satisfying $\Phi$ leads to another state satisfying $\Phi$).

In this paper we describe an algorithm for the automatic construction
of inductive invariants for array-based systems
(Section~\ref{sec:abs}). This algorithm, based on both IC3
\cite{bradley} and Lazy Abstraction \cite{mcmillan}, builds an
inductive invariant by unwinding a graph (Section~\ref{sec:unwinding})
building a forward abstract reachability of our system. This unwinding
is described as a set of non deterministic rules. We then provide an
implementation in Cubicle \cite{thesemebsout},
\cite{cav2012}, \cite{fmcad2013} (Section~\ref{sec:implementation}) of these
rules and test its effectiveness on several cache coherence protocols
(Section~\ref{sec:benchmarks}).

\section{Array-based Systems}
\label{sec:abs}

An array-based system is described in \cite{mcmt-foundations} as
first-order logic formulas on arrays. Such a system can be described
as a set of basic types, a set $X$ of \textit{system variables}
associated to type (built as usual with basic types and standard
constructions), a formula \textit{init} representing the initial
states and a set $\Delta{}$ of transition rules $\tau{}_i(X, X')$
($X'$ is the set $X$ where all the variables are primed which
represents the \emph{next state} reached after the application of a
transition). Since we work on \emph{parameterized} programs, our
arrays are indexed by an infinite type \textit{proc}.

We describe the Dekker mutual exclusion algorithm
as an array-based system. Each process has two boolean variables,
\textit{want} (stating that the process wants to enter in critical
section or not) and \textit{crit} (stating that the process is in
critical section or not). There is a global variable \textit{turn} of
type \textit{proc} that tracks which process can go into the critical
section. Since we work in the array-based systems fragment, we
represent the local variables as arrays indexed by processes and
containing booleans. The set $X$ contains two arrays,
\textit{want[proc] : bool} and \textit{crit[proc] : bool} and the
global variable \textit{turn : proc}. Initially, no process is or
wants to be in critical section. Three transitions can be triggered,
one to require an access to the critical section, one to enter in it
and one to exit it. According to the previous description, we 
write this algorithm as in Figure~\ref{fig:dekker_abs}.

\begin{figure}[h]
  \scalebox{.8}{
    \vbox{%
      \centering
      \[
        \begin{array}{ll}
          \mathtt{turn} : proc
          & \\
          \mathtt{crit}[proc] : bool
          & \\
          \mathtt{want}[proc] : bool
          & \\
          & \\
          init: \forall p. \hspace*{-0.5em}
          &\neg \mathtt{want}[p] ~\land~ \neg \mathtt{crit}[p]\\[1em]
          \multirow{2}{*}{$req: \exists p.$}  \hspace*{-0.5em}
          & \neg \mathtt{want}[p] \\
          & \mathtt{want}'[p] \\[1em]
          \multirow{2}{*}{$enter: \exists p.$}  \hspace*{-0.5em}
          & \mathtt{want}[p] ~\land~ \mathtt{turn} = p \\
          &\mathtt{crit}'[p] \\[1em]
          \multirow{2}{*}{$exit:\exists p_1,~ p_2.$}  \hspace*{-0.5em}
          & \mathtt{crit}[p_1] \\
          & \neg\mathtt{want}'[p_1] \land \neg\mathtt{crit}'[p_1]\\
          &        \land \mathtt{turn}' = p_2 \\[1em]
        \end{array}
      \]
    }
  }
  \caption{Dekker algorithm as array-based system}
  \label{fig:dekker_abs}
\end{figure}

Since we focus on safety problems (\textit{nothing bad happens}),
we need to define what is considered as \emph{bad states}. In that
case these states would be defined with the following formula :\\[-2.1em]

\[unsafe \equiv \exists p_1,~ p_2.~ p_1 \neq p_2 ~\land~
  \mathtt{crit}[p_1] ~\land~ \mathtt{crit}[p_2]\]

Our goal is then to prove that no state represented by $unsafe$ is
reachable from $init$ (which can be seen as : there exists no path
$init = X_0 \xrightarrow{p_1} X_1 \xrightarrow{\dots} \dots
\xrightarrow{p_n} X_n = unsafe$
with $p_i \in \{req, enter, exit\}$).  To do so on parameterized
system, one of the main algorithm came from Ghilardi et
al. with MCMT \cite{ghilardiMCMT} and builds the set of all
reachable states by \emph{backward reachability} (\textit{starting,
  then, from the unsafe state}) and checks if this set contains an
initial state. In this paper, we implement a different algorithm
which offers a wider range of possibilities in terms of reachabilty
construction.
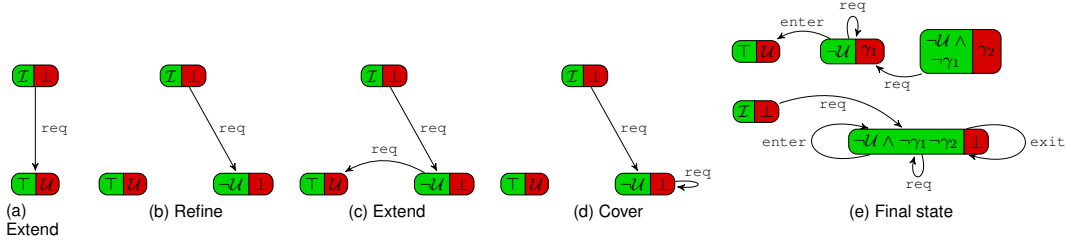
\begin{figure*}[!t] \scriptsize
  \centering
  \scalebox{0.8}{
    \vbox{
      \begin{subfloat}[Extend]{
          \begin{tikzpicture}[]
            \node (BadN) at (0,0) [noeudex]
            {
              $\top$
              \nodepart{second}$\unsafe$
            };
            \node (root) at (0,1.8) [noeudex]
            {
              $\init$
              \nodepart{second}$\bot$
            };
            \path[myarrow] (root) edge node[right]{\te{req}} (BadN);
          \end{tikzpicture}
        }
      \end{subfloat}
      \hfill
      \begin{subfloat}[Refine]{
          \begin{tikzpicture}[]
            \node (BadN) at (-1,0) [noeudex]
            {
              $\top$
              \nodepart{second}$\unsafe$
            };
            \node (root) at (0,1.8) [noeudex]
            {
              $\init$
              \nodepart{second}$\bot$
            };
            \node (node1) at (1,0) [noeudex]
            {
              $\neg \unsafe$
              \nodepart{second}$\bot$
            };
            
            \path[myarrow] (root) edge node[right]{\te{req}} (node1);
          \end{tikzpicture}
        }
      \end{subfloat}
      \hfill
      \begin{subfloat}[Extend]{
          \begin{tikzpicture}[]
            \node (BadN) at (-1,0) [noeudex]
            {
              $\top$
              \nodepart{second}$\unsafe$
            };
            \node (root) at (0,1.8) [noeudex]
            {
              $\init$
              \nodepart{second}$\bot$
            };
            \node (node1) at (1,0) [noeudex]
            {
              $\neg \unsafe$
              \nodepart{second}$\bot$
            };
            
            \path[myarrow] (root) edge node[right]{\te{req}} (node1);
            \path[myarrow] (node1) edge[bend right] node[above]{\te{req}} (BadN);

          \end{tikzpicture}
        }
      \end{subfloat}
      \hfill
      \begin{subfloat}[Cover]{
          \begin{tikzpicture}[]
            \node (BadN) at (-1,0) [noeudex]
            {
              $\top$
              \nodepart{second}$\unsafe$
            };
            \node (root) at (0,1.8) [noeudex]
            {
              $\init$
              \nodepart{second}$\bot$
            };
            \node (node1) at (1,0) [noeudex]
            {
              $\neg \unsafe$
              \nodepart{second}$\bot$
            };
            
            \path[myarrow] (root) edge node[right]{\te{req}} (node1);
            \path[myarrow] (node1) edge[out=5, in=-5, loop] node[above]{\te{req}} (node1);

          \end{tikzpicture}
        }
      \end{subfloat}
      \begin{subfloat}[Final state]{
          \begin{tikzpicture}[]
            \node (BadN) at (-0.8,0) [noeudex]
            {
              $\top$
              \nodepart{second}$\unsafe$
            };
            \node (root) at (-0.8,-1) [noeudex]
            {
              $\init$
              \nodepart{second}$\bot$
            };
            \node (node1) at (0.8,0) [noeudex]
            {
              $\neg \unsafe$
              \nodepart{second}$\badcube_1$
            };
            \node (node2) at (2.6,0) [noeudex]
            {
              \nodepart[text width=0.7cm]{one}$\neg \unsafe \land \neg \badcube_1$
              \nodepart{second}$\badcube_2$
            };
            \node (node3) at (1.9,-1.5) [noeudex]
            {
              \nodepart[]{one}$\neg \unsafe \land \neg \badcube_1 \neg \badcube_2$
              \nodepart{second}$\bot$
            };

            \path[myarrow] (root) edge[bend left] node[pos=0.4,below]{\te{req}} (node3);
            \path[myarrow] (node1) edge[loop above] node[above]{\te{req}} (node1);
            \path[myarrow] (node1) edge[bend right] node[above]{\te{enter}} (BadN);
            \path[myarrow] (node2) edge[bend left] node[below]{\te{req}} (node1);
            \path[myarrow] (node3) edge[loop below] node[below]{\te{req}} (node3);
            \path[myarrow] (node3) edge[loop left] node[left]{\te{enter}} (node3);
            \path[myarrow] (node3) edge[loop right] node[right]{\te{exit}} (node3);
          \end{tikzpicture}
        }
      \end{subfloat}
    }
  }
  \caption{First four steps of the unwinding and final state of the
    graph (\textit{sink rules are not shown)}}
  \label{fig:dekker_run}
\end{figure*}

\section{Program unwinding}
\label{sec:unwinding}

This algorithm starts also from the $unsafe$ formula but tries to
build an invariant of the system that does not contain it. Before
going into details, we give a brief explanation. This invariant is
iteratively built as an inductive invariant $\Theta$ that does not
contain $unsafe$ :

\begin{itemize}
\item if $\Theta ~\land~ \Delta ~\land~ \neg\Theta'$ is
  unsatisfiable then we found an inductive invariant
\item if $\Theta ~\land~ \Delta ~\land~ \neg\Theta'$ is
  satisfiable, our candidate invariant is not inductive and we try to
  refine it until we either discover that there is no such refinement
  or we find some.
\end{itemize}

\noindent
For Dekker's algorithm, for example, let's take $\Theta = \neg unsafe
= \forall p_1\neq p_2. \neg crit[p_1] \lor \neg crit[p_2]$
:
\begin{itemize}
\item $\Theta ~\land~ \Delta ~\land~ \neg\Theta'$ is satisfiable (if
  we take, for example, the following state : $\varphi_1 = crit[p_1] ~\land~
  want[p_2] ~\land~ turn = p_2 ~\land~ \neg crit[p_2]$, $\varphi_1
  \models \Theta$ but if we apply $enter$ to it we obtain the state
  $\varphi_2 = crit[p_1] ~\land~ crit[p_2] ~\land~ \dots$ and $\varphi_2
  \not\models \Theta$.)
\item We need to create $\Theta' = \Theta ~\land~ \rho$ which is a
  refinement of $\Theta$ that does not contain $\varphi_1$.
\end{itemize}

\noindent
To do so, we build an \emph{unwinding} of the algorithm as a quadruple
$\langle V, E, \mathcal{W}, \mathcal{B} \rangle$, where:
\begin{itemize}
\item $\langle V, E\rangle$ is a rooted graph with edges
  labeled by transitions from $\Delta$;
\item $\mathcal{W}$ associates a formula (called \emph{world
    of the vertex}) to each vertex;
\item $\mathcal{B}$ associates a formula (called \emph{bad
    part of the vertex}) to each vertex.
\end{itemize}
This graph contains three initial vertices :
\begin{enumerate}
\item[$\epsilon$] : the root vertex, $\world(\epsilon) = init$ and
  $\bad(\epsilon) = \bot$;
\item[$\beta$] : the unsafe vertex, $\world(\beta) = \top$ and
  $\bad(\beta) = unsafe$;
\item[$\omega$] : the sink vertex, $\world(\omega) = \bot$ and
  $\bad(\omega) = \bot$.
\end{enumerate}

We call $V^\epsilon = \{v \in V,~ \epsilon \xrightarrow{*}v \in E\}$
(\textit{i.e.} the set of vertices that are linked to the
root). $\world(v) \models_\tau \world(v') \equiv \world(v)
~\land~ \tau \models \world(v')$
\\[1em]
\indent The idea behind this unwinding it that if we manage to create a graph
$\mathcal{G}$ of a system $\mathcal{S} =\langle init, \Delta \rangle$
where every vertex in $V^\epsilon$ does not contain a bad part and
from which no more transition can be taken, then the disjunction of
their worlds ($\Theta = \bigvee_{v \in V^\epsilon}\world(v)$) is an
invariant of the system ($init \models \Theta$ and $\Theta
\models_\Delta \Theta$).




We now propose a set of non-deterministic rules for building this
unwinding.  Let $\langle X, init, \Delta, unsafe\rangle$ be an
array-based system. Initially, $\mathcal{G}$ is defined as follow :
\begin{enumerate}
\item[-] $V = \{\epsilon, \omega, \beta\}$
\item[-] $E = \emptyset$
\end{enumerate}

The unwinding works by the non-deterministic application of the
following rules :

\begin{myrule}[\textbf{Extend}]
  If $\exists v \in V, \tau \in \Delta.~\world(v) \models_\tau \top$
  and $\nexists v'. v \xrightarrow{\tau} v' \in E$ then $E =
  E \cup \{v \xrightarrow{\tau} \beta\}$
\end{myrule}

\begin{myrule}[\textbf{Refine}]
  If $\exists v, v' \in V, \tau \in \Delta.~ v \xrightarrow{\tau} v'
  \in E$, $\bad(v') \neq \bot$, $\exists \varphi. \world(v)
  \models_\tau \varphi$ and $\varphi \models \neg \bad(v')$ then we
  create a new vertex $v''$ such that $\world(v'') = \world(v')
  ~\land~ \varphi$ and $E = E \cup \{v \xrightarrow{\tau} v''\}
  \setminus \{v \xrightarrow{\tau} v'\}$
\end{myrule}

\begin{myrule}[\textbf{Propagate}]
  If $\exists v, v' \in V, \tau \in \Delta.~ v \xrightarrow{\tau} v'
  \in E$, $\bad(v') \neq \bot$, $\exists \gamma.~ \gamma \models
  \world(v)$, and $\gamma \models_\tau \bad(v')$ then  $B(v)
  \leftarrow \gamma$
\end{myrule}

\begin{myrule}[\textbf{Cover}]
  If $\exists v, v' \in V, \tau \in \Delta.~ v \xrightarrow{\tau} v'
  \in E$, $v'' \in V$ such that $\world(v'') \models \world(v')$ and
  $\world(v) \models_\tau \world(v'')$ then $E = E \cup \{v
  \xrightarrow{\tau} v''\} \setminus \{v \xrightarrow{\tau} v'\}$
\end{myrule}

\begin{myrule}[\textbf{Sink}]
  If $\exists v \in V, \tau \in \Delta.~\world(v) \models_\tau \bot$
  and $\nexists v'. v \xrightarrow{\tau} v' \in E$ then $E =
  E \cup \{v \xrightarrow{\tau} \omega \}$
\end{myrule}

$S$ is safe if and only if no rule can be applied to $\mathcal{G}$, an
unwinding $\mathcal{S}$ and $\bad(\epsilon) = \bot$. Intuitively,
since no more transitions can be taken and all the vertices connected
to the root are not bad, root will never be able to lead to unsafe.

\section{Example}
\label{sec:example}

The example shown in Figure~\ref{fig:dekker_run} describes the first
four runs of the unwinding on the Dekker's algorithm (we decided not
to show $\omega$ since it just serves as a sink for the transitions
that can not be taken from a vertex) :

\begin{itemize}
\item[(a)] initially, the only rule that can be applied is the
  \textbf{Extend} rule from $\epsilon$ (with $\world(\epsilon) \equiv
  init \equiv \forall p.~ \neg \mathtt{want}[p] ~\land~ \neg
  \mathtt{crit}[p]$) with $req$;
\item[(b)] we can only apply, then, the \textbf{Refine} rule because
  $init \nvDash_{req} unsafe \equiv \bad(\beta)$. We create a new vertex
  called $v_1$;
\item[(c)] we can chose, here, to apply the \textbf{Extend} rule from
  the new vertex with any transition. We chose to take the transition $req$;
\item[(d)] $\world(v_1) \nvDash_{req} \bad(\beta)$ and $\world(v_1)
  \models_{req} \world(v_1)$ so we can apply the cover rule;
\item[(e)] if we keep applying these rules, we reach a fixpoint at
  the third created vertex.
\end{itemize}

\section{Implementation}
\label{sec:implementation}

We implemented this unwinding in Cubicle
\footnote{\url{cubicle.lri.fr/far}}. To do so we had to chose a
deterministic strategy depending on multiple parameters :
\begin{itemize}
\item the order in which the rules are applied;
\item which vertex and transition should be taken for the
  \textbf{Extend} rule;
\item which formula $\varphi$ should we take for the \textbf{Refine}
  rule;
\item which formula $\gamma$ should we take for the \textbf{Propagate}
  rule;
\item which vertex $v''$ should we take for the \textbf{Cover} rule.
\end{itemize}

Based on this problems, we came out with the following algorithm (we
write $v = (W, B)$ to denote the fact that $\world(v) = W$ and
$\bad(v) = B$) :

\begin{algorithm}
  \caption{Graph unwinding - main loop}\label{alg:main_loop}
  \begin{algorithmic}[]
    \Procedure{far-cubicle}{$\mathcal{S} = \langle init, \Delta, unsafe\rangle$}
    \State $\epsilon \gets (init, \bot)$
    \State $\beta \gets (\top, unsafe)$
    \State $\omega \gets (\bot, \bot)$
    \State $V\gets \{\epsilon, \beta, \omega\}$
    \State $E = \emptyset$
    \State \Call{push}{$\mathcal{Q}$, $\epsilon$}
    \Comment{$\mathcal{Q}$ is a priority queue}
    \While{\Call{not\_empty}{$\mathcal{Q}$}}
    \State $v\gets$\Call{pop}{$\mathcal{Q}$}
    \ForAll{$\tau \in \Delta$}
    \If{$\world(v) \models_\tau \top$}
    \State $E = E \cup \{v \xrightarrow{\tau} \beta\}$
    \State \Call{unwind}{$v \xrightarrow{\tau} \beta$}
    \Else $~E = E \cup \{v \xrightarrow{\tau} \omega\}$
    \EndIf
    \EndFor
    \EndWhile
    \State \Return \textcolor{green}{\textbf{safe}}
    \EndProcedure
  \end{algorithmic}
\end{algorithm}

\begin{algorithm}
  \caption{Graph unwinding - unwinding procedure}\label{alg:main_loop}
  \begin{algorithmic}[]
    \Procedure{unwind}{$v \xrightarrow{\tau} v'$}
    \If{$\bad(v) = \bot ~\land~ \bad(v') \neq \bot$}
    \Switch{\Call{close}{$v \xrightarrow{\tau} v'$}}
    \Case{Covered v''}
    \State $E = E \cup \{v \xrightarrow{\tau} v''\} \setminus \{v
    \xrightarrow{\tau} v'\}$
    \State \Call{unwind}{$v \xrightarrow{\tau} v''$}
    \EndCase
    \Case{Bad $\varphi$}
    \If{$v = \epsilon$} \Return \textcolor{red}{\textbf{unsafe}}
    \Else
    \State $\bad(v) \gets \varphi$
    \ForAll{$u \xrightarrow{\tau'} v$}
    \State \Call{unwind}{$u \xrightarrow{\tau} v$}
    \EndFor
    \EndIf
    \EndCase
    \Case{Refined v''}
    \State $E = E \cup \{v \xrightarrow{\tau} v''\} \setminus \{v
    \xrightarrow{\tau} v'\}$
    \State \Call{push}{$\mathcal{Q}, v''$}
    \EndCase
    \EndSwitch
    \EndIf
    \EndProcedure
  \end{algorithmic}
\end{algorithm}

This algorithm picks a vertex $v$ from a priority queue (which initially
contains only the root vertex) and for all the transitions, adds an edge
to the graph from this transition to the sink vertex if the formula
represented by $v$ is inconsistent with the transition or to
the unsafe vertex if the transition can be taken. If the edge goes to
a vertex $v'$ that is not the sink, the procedure \textsc{unwind} is called
on it. This procedure checks if $\mathcal{B}(v) \neq \bot$ or if
the $\mathcal{B}(v') = \bot$ and if both these conditions are false it
tries to \emph{close} the edge. An edge is \emph{closed} if :
\begin{itemize}
\item $\world(v) \models_\tau \mathcal{B}(v')$. In this case, all
  the edges coming to it must be unwinded again;
\item there exists another vertex $v''$ such that $v \models_\tau v''$
  and $\mathcal{v''} \models \mathcal{W}(v')$. In this case, the edge
  from $v$ to $v'$ is deleted and a new one from $v$ to $v''$ is
  created and unwinded;
\item $\world(v) \not\models_\tau \bad(v')$. A counter example $\varphi$ is
  found a new node $v''$ is created with $\world(v'') \equiv
  \world(v') \land \varphi$ and pushed in the queue.
\end{itemize}
If all the edges are closed and the queue is empty, the system is
safe. If the propagation of bad parts reaches the root vertex, the
system is unsafe.

\begin{algorithm}
  \caption{Graph unwinding - closing edge procedure}\label{alg:main_loop}
  \begin{algorithmic}[1]
    \Procedure{close}{$v \xrightarrow{\tau} v'$}
    \If{$\exists v''.~ \world(v'') \models \world(v') ~\land~
      \world(v) \models_\tau \world(v'')$}
    \State\Return Covered $v''$
    \ElsIf{$\world(v) \models_\tau \bad(v')$}
    \State\Return Bad \Call{pre}{$\bad(v'), \tau$}\label{lbl:pre_bad}
    \Else
    \State $v'' \gets (\world(v') ~\land~$ \Call{generalize}{$\neg \bad(v')}, \bot)$\label{lbl:neg_refine}
    \State\Return Refined v''
    \EndIf
    \EndProcedure
  \end{algorithmic}
\end{algorithm}

As we can see on line~\ref{lbl:pre_bad} of the \textsc{close}
procedure, the formula $\gamma$ chosen for the \textbf{Propagate} rule
is the pre image of the bad formula of the vertex $v'$.
Also, on line~\ref{lbl:neg_refine} of the \textsc{close}
procedure, the formula $\varphi$ chosen for the \textbf{Refine} rule
is a generalization of the negation of the bad part of the vertex
$v'$. These are, of course,
implementation choices. Other implementation could involve \emph{model
  finding}, \emph{interpolants} ... In our case, the generalization is
a naive one consisting in taking the smallest part of the
resulting formula that was not already taken and that still satifies
the conditions of the \textbf{Refine} rule.
\section{Benchmarks}
\label{sec:benchmarks}

We compared our implementation to the backward reachability algorithm
already implemented in Cubicle (without the invariants inference
implemented with BRAB \cite{thesemebsout}, \cite{fmcad2013}) and
obtained the following results (the
timeout was set to 5 minutes and the $\alpha$ version uses an
abstraction engine related to the approximation implemented in BRAB to
get better refinement):\\[1em]

\begin{tabular}[h]{|c|c|c|c|}
  Protocol & Cubicle & FAR & FAR-$\alpha$ \\
  \hline
  \texttt{dekker}& 0.04s & 0.04s & 0.03s\\
  \texttt{mux\_sem} & 0.04s & 0.05s & 0.03s \\
  \texttt{german}-ish & 0.06s & 0.1s & 0.55s\\
  \texttt{german}-ish2 & 0.13s & 0.11s & 0.65s\\
  \texttt{german}-ish3 & 1.2s & 8.3s & 0.65s \\
  \texttt{german}-ish4 & 3.5s & 2.5s & 0.75s\\
  \texttt{german}-ish5 & 1.9s & 8.2s & 0.60s\\
  \texttt{german} & 18s & 5.8s & 4.25s \\
  \texttt{szymanski\_at} & TO & 13s & 2.60s \\
  \texttt{szymanski\_na} & TO & TO & 16s 
\end{tabular}

\vspace*{1em}As we can see in this table, this algorithm is
competitive and even better when good refinements can be found.

\section{Related Works}
There has been a lot of research in software model checking and
Property-Driven Reachability. This type of algorithm was first
introduced by Bradley in \cite{bradley} and McMillan revisited his
Lazy Annotation (which shares similarities with PDR algorithms) in
\cite{mcmillan_revisited} or the recent approach from Cimatti et
al. \cite{cimattiIC3} and Z3 with a PDR approach in
\cite{DBLP:conf/sat/HoderB12} and \cite{DBLP:conf/cav/HoderBM11}.
Even though some of these tools are supposed to work on parameterized
systems, we were either not able to find them or they were not able to
prove our examples.

\section{Conclusion}

We presented the problem of parameterized protocol verification and
gave an algorithm to automatically do it. This new algorithm was
implemented in Cubicle and successfully applied to many
cache coherence protocols.

This algorithm could be improved with a better generalisation engine
(allowing to explore less vertices), an incremental approach (the
parameterized aspect of our language makes it hard to
\textit{remember} the state of our SMT solver). Other optimizations
could involve a novel way of refining our formulas (it is clear that
the best refinements are inductive invariants but it is still an open
problem as how to find these).

Some optimizations were not documented in this article such as
\begin{itemize}
\item \emph{Set-theoretic test} : some formulas are trivially
  unsatisfiable and don't require call to the SMT solver;
\item \emph{relevant instantiations} : handling universally quantified
  formulas can lead to multiple useless instantiations that are
  trivially unsatisfiable or valid and do not help the SMT solver to
  solve the whole formula. This optimization allows to gain a
  significant time in the SMT solver.
\item \emph{selecting good bads} : handling bad parts from the ones
  with less processes involved allows to control the number of
  processes that have to be instantiated when checking the
  satisfiability of formulas. It is mandatory, if we want to have a
  competitive algorithm, that we handle the bad parts cleverly (this
  can be done in the priority queue).
\end{itemize}

\bibliographystyle{IEEEtran}
\bibliography{fmcad2017biblio}
      %




\end{document}